\documentclass[aip,jmp,amsmath,amssymb,reprint]{revtex4-1}
\usepackage{graphicx}
\usepackage{dcolumn}
\usepackage{bm}
\usepackage{multirow}
\usepackage{amssymb}
\usepackage{amsmath}
\usepackage{float}
\restylefloat{table}

\usepackage{placeins}
\usepackage{afterpage}
\usepackage{subfigure}
\usepackage[colorlinks=true,
            linkcolor=blue,
            urlcolor=blue,
            citecolor=blue,
            breaklinks=true]{hyperref}
\begin{document}
\preprint{AIP/123-QED}

\title{Frequency power spectra of global quantities in magnetoconvection}

\author{Sandip Das}
\author{Krishna Kumar}
\email{kumar.phy.iitkgp@gmail.com}
\affiliation{Department of Physics, Indian Institute of Technology Kharagpur, Kharagpur-721302, India}

\date{\today}
\begin{abstract}
We present the results of direct numerical simulations  of  power spectral densities for kinetic energy, convective entropy and heat flux for unsteady Rayleigh-B\'{e}nard magnetoconvection in the frequency space. For larger values of frequency, the power spectral densities for all the global quantities vary with frequency $f$ as  $f^{-2}$. The scaling exponent is independent of Rayleigh number, Chandrasekhar's number  and thermal Prandtl number.
\end{abstract}

\maketitle

\section{\label{sec:Intro}Introduction}

The temporal fluctuations of spatially averaged (or, global) quantities are of interest in several fields of research including turbulent flows~\cite{Fauve_etal_1993,Niemela_etal_2000,Aumaitre_Fauve_epl_2003,Gerolymos_etal_2014}, nanofluids~\cite{Kakac_Pramuanjaroenkij_2009}, biological fluids~\cite{Tsimring_2014,Deco_etal_2017},  geophysics~\cite{Damon_etal_1978,Tyler_etal_2017}, 
phase transitions~\cite{Mobilia_etal_2007,Brennecke_etal_2013}. The probability density function (PDF) of the temporal fluctuations of thermal flux in turbulent Rayleigh-B\'{e}nard convection (RBC) was found to have normal distribution with slight asymmetries at the tails. The direct numerical simulations (DNS) of the Nusselt number $\mathrm{Nu}$, which is a measure of thermal flux,  also showed the similar behaviour in presence of the Lorentz force~\cite{Das_Kumar_2019}.
The power spectral density (PSD) of the thermal flux in the frequency ($f$) space~\cite{Aumaitre_Fauve_epl_2003,Das_Kumar_2019,Hirdesh_etal_2014} was found to vary as  $f^{-2}$.  In this work, we present the results obtained by DNS of temporal signals of global quantities: spatially averaged kinetic energy per unit mass $E $, convective entropy per unit mass $E_{\Theta}$ and Nusselt number $\mathrm{Nu}$ in unsteady Rayleigh-B\'{e}nard magnetoconvection (RBM)~\cite{Chandrasekhar_1961,Fauve_etal_1984,Weiss_Proctor_2014}. The kinetic energy as well as the entropy vary with frequency as $f^{-2}$ at relatively higher frequencies. In this scaling regime, the scaling exponent does not depend on the Rayleigh number $\mathrm{Ra}$, Prandtl number $\mathrm{Pr}$ and Chandrasekhar's number $\mathrm{Q}$.  

\section{\label{sec:system} Governing equations}
The physical system consists of a thin layer of a Boussinesq fluid (e.g., liquid metals, melt of some alloys (i.e., $NaNO_3$ melt), nanofluids, etc.) of density $\rho_0$ and electrical conductivity $\sigma$ confined between two horizontal plates, which are made of electrically non-conducting but thermally conducting materials. The lower plate is heated uniformly and the upper plate is cooled uniformly so that an adverse temperature gradient $\beta$ is maintained across the fluid layer. A uniform magnetic field $B_0$ is applied in the vertical direction. The positive direction of the $z$- axis is in the direction opposite to that of the acceleration due to gravity $g$.  The basic state is  the conduction state with no fluid motion. The stratification of the steady temperature field $T_s (z)$, fluid density $\rho_s(z)$ and pressure field $P_s(z)$, in the conduction state~\cite{Chandrasekhar_1961}, are given as:
\begin{eqnarray}
T_s (z) &=& T_b + \beta z,\\
\rho_s (z) &=& \rho_0 \left[1 + \alpha \left(T_b - T_s (z)\right)\right],\\
P_s (z) &=& P_0 - \left[\rho_0 g \left(z + \frac{1}{2}\alpha \beta z^2 \right) + \frac{{B_{0}}^2}{8\mu_0 \pi}\right],
\end{eqnarray}
where $T_b$ and $\rho_0$ are temperature and  density of the fluid at the lower plate, respectively.  $P_0$ is a constant pressure in the fluid and $\mu_0$ is the permeability of the free space.

As soon as the temperature gradient across the fluid layer is raised above a critical value $\beta_c$ for fixed values of all fluid parameters (kinematic viscosity $\nu$, thermal diffusivity $\kappa$, thermal expansion coefficient $\alpha$) and the externally imposed magnetic field $B_0$, the convection sets in. All the fields are perturbed due to convection and they may be expressed as:

\begin{eqnarray}
\rho_s (z) \rightarrow \tilde{\rho}(x, y, z, t) &=& \rho_s (z) + \delta \rho (x, y, z, t),\\
T_s (z) \rightarrow T(x, y, z, t) &=& T_s (z) + \theta (x, y, z, t),\\
P_s (z) \rightarrow P(x, y, z, t) &=& P_s (z) + p (x, y, z, t),\\
{\bm{\mathrm{B}}}_0 \rightarrow {\bm{\mathrm{B}}} (x, y, z, t) &=& {\bm{\mathrm{B}}}_0 + \bm{\mathrm{b}} (x, y, z, t),
\end{eqnarray}

where $\bm{\mathrm{v}}(x,y,z,t)$, $\mathrm{p}(x,y,x,t)$, $\theta (x,y,z,t)$ and $\bm{\mathrm{b}}(x,y,z,t)$ are the fluid velocity, perturbation in the fluid pressure  and the convective temperature and the induced magnetic field, respectively, due to convective flow.  The perturbative fields are made dimensionless by measuring all the length scales in units of the clearance $d$ between two horizontal plates, which is also the thickness of the fluid layer.  The time is measured in units of  the free fall time  $\tau_{f} = 1/\sqrt{\alpha  \beta g}$. The convective temperature field $\theta$ and the induced magnetic field $\bf{b}$ are dimensionless by $\beta d$ and $B_0 \mathrm{Pm}$, respectively. The magnetoconvective dynamics is then described by the following dimensionless equations:

\begin{eqnarray}
& D_t\bm{\mathrm{v}}=-\nabla p+\sqrt{\frac{\mathrm{Pr}}{\mathrm{Ra}}}\nabla^2\bm{\mathrm{v}}+\frac{\mathrm{Q}\mathrm{Pr}}{\mathrm{Ra}}\partial_z\bm{\mathrm{b}}+\theta\bm{\mathrm{e}}_3,\label{eq:mom-v}\\
&\nabla^2\bm{\mathrm{b}} = -\sqrt{\frac{\mathrm{Ra}}{\mathrm{Pr}}}\partial_z\bm{\mathrm{v}}, \label{eq:mag-v}\\
& {D_t \theta} = \sqrt{\frac{1}{\mathrm{Ra} \mathrm{Pr}}}\nabla^2\theta + {\mathrm{v}}_3, \label{eq:theta}\\
&\nabla\cdot\bm{\mathrm{v}} = \nabla\cdot\bm{\mathrm{b}}=0,\label{eq:cont}
\end{eqnarray}
where $D_t \equiv \partial_t + (\bm{\mathrm{v}}\cdot\nabla)$ is the material derivative. As the magnetic Prandtl number $\mathrm{Pm}$ is very small ($\le 10^{-5}$) for all terrestrial fluids, we set $\mathrm{Pm}$ equal to zero in the above. The induced magnetic field is then slaved to the velocity field. We consider the idealized boundary ({\sl stress-free}) conditions for the velocity field on the horizontal boundaries. The relevant boundary conditions~\cite{Chandrasekhar_1961,Basak_etal_pre2014} at horizontal plates, which are located at $\mathrm{z} = 0$ and $ \mathrm{z}= 1$, are: 

\begin{equation}
\frac{\partial \mathrm{v}_1}{\partial z}=\frac{\partial \mathrm{v}_2}{\partial z}=\mathrm{v}_3=\mathrm{b}_1=\mathrm{b}_2= \frac{\partial \mathrm{b}_3}{\partial z} =\theta=0. \label{eq:bound-v} 
\end{equation}
All fields are considered periodic in the horizontal plane. The dynamics of the flow (as $\mathrm{Pm} \rightarrow 0$) is controlled by three dimensionless parameters: (1) Rayleigh number $\mathrm{Ra} = \frac{\alpha \beta g d^4}{\nu \kappa}$, (2) Prandtl number $\mathrm{Pr} = \frac{\nu}{\kappa}$ and (3) Chandrasekhar's number $\mathrm{Q}=\frac{\sigma B_0^2 d^2}{\rho_0 \nu}$. 
The critical values of Rayleigh number $\mathrm{Ra_c}$ and the critical wave number $k_c$ are~\cite{Chandrasekhar_1961}:  
\begin{eqnarray}
& \mathrm{Ra}_{c}(\mathrm{Q}) = \frac{\pi^2 + k_{c}^2}{k_{c}^2}\big[ ( \pi^2 + k_{c}^2 )^{2} + \pi^{2}\mathrm{Q} \big],\label{eq:Ra}\\
& k_{c}(\mathrm{Q}) = \pi \sqrt{a_{+} + a_{-} - \frac{1}{2}},\label{eq:k}
\end{eqnarray}
where

\begin{equation}
 a_{\pm} = \Bigg( \frac{1}{4} \Big[\frac{1}{2} + \frac{\mathrm{Q}}{\pi^{2}} \pm \big[ \big( \frac{1}{2} + \frac{\mathrm{Q}}{\pi^{2}} \big)^{2} - \frac{1}{4} \big]^{\frac{1}{2}}\Big] \Bigg)^{\frac{1}{3}}. \label{eq:a} 
 \end{equation}
 The kinetic energy $E$ and convective entropy $E_{\Theta}$ per unit is mass are defined as: $E = \frac{1}{2} \int{\mathrm{v}^2 dV}$ and $E_{\Theta} = \frac{1}{2} \int{\theta^2 dV}$, respectively. The Nusselt number $\mathrm{Nu}$ , which is the ratio of total heat flux and the conductive heat flux across the fluid layer, is defined as: $\mathrm{Nu} = 1 + \frac{\sqrt{\mathrm{Ra} \mathrm{Pr}}}{V}\int{\mathrm{v}_3\theta dV}$. 
 
The system of equations may also be useful for investigating magnetoconvection in nanofluids with low concentration non-magnetic metallic nanoparticles~\cite{Das_Kumar_2019}. A homogeneous suspension of nanoparticles in a viscous fluid works as a nanofluid. As the fluid properties depend on the base fluid and the nano-particles, their effective values may be used for the nanofluid. All fluid parameters are may be replaced by their effective values in the presence of nanoparticles in a simple model.  If $\phi$ is the volume fraction of the spherically shaped nanoparticles, the effective form of the  density and electrical conductivity of the nanofluid may be expressed as:

\begin{eqnarray}
\rho &=& (1-\phi) \rho_f + \phi \rho_p,\\
\sigma &=& (1-\phi) \sigma_f + \phi \sigma_p,
\end{eqnarray}
where $\rho_f$ and $\sigma_f$ are the density and electrical conductivity of the base fluid, respectively. Here $\rho_p$ is the density and $\sigma_p$ is the electrical conductivity of the nanoparticles. The effective thermal conductivity $K$~\cite{Maxwell_1873} is expressed as:

\begin{equation}
K = K_{f} \left[ \frac{(K_{p}+2K_{f}) - 2\phi(K_{f}-K_{p})}{(K_{p} + 2K_{f}) + \phi(K_{f}-K_{p})} \right],
\end{equation}
where $K_f$ and  $K_p$ are  the thermal conductivity of the base fluid and that
of the spherical shaped nanoparticles, respectively. Similarly, the effective specific ccapacity $c_V$ may be expressed through the following relation~\cite{Selimefendigil_Oztop_2014}:

\begin{equation}
(\rho c_V) = (1-\phi)(\rho c_V)_{f} + \phi (\rho c_V)_{p}.
\end{equation} 

The effective dynamic viscosity $\mu$ of the nanofluid~\cite{Brinkman_1952} may also be expressed as:

\begin{equation}
\mu = \mu_{f}(1-\phi)^{-2.5}.
\end{equation}
The relevant values of effective fluid parameters may be used in the set of equations \ref{eq:mom-v}-\ref{eq:cont} for investigating flow properties in nanofluids.

\begin{figure}[ht]
	\centering
	\includegraphics[height=!,width=9.0 cm]{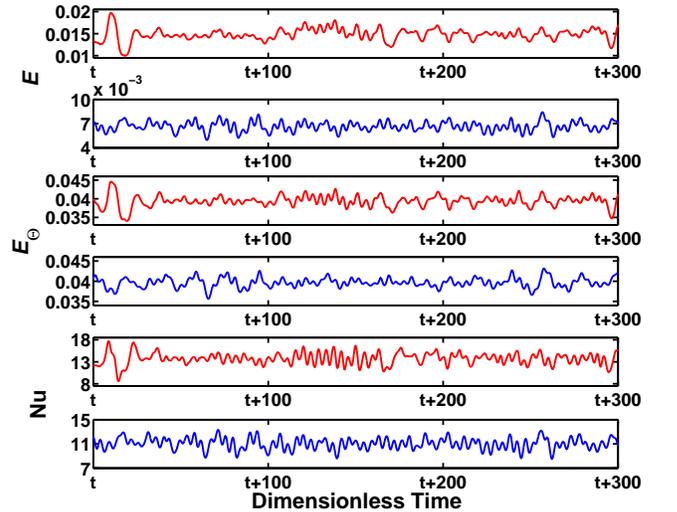}
	\caption{\label{Time_sig} (Colour online)
		Temporal variations of the kinetic energy $E$, entropy $E_{\Theta}$ and Nusselt number $\mathrm{Nu}$ for Rayleigh number $\mathrm{Ra} = 5.0\times10^5$ and Prandtl number $\mathrm{Pr}=4.0$. The light gray (red) curves are for Chandrasekhar number $\mathrm{Q}= 100$ and the gray (blue) curves are for $\mathrm{Q}= 100$. }  
\end{figure}

\begin{figure*}[htp]
	\centering
	\includegraphics[height=!,width=18.0 cm]{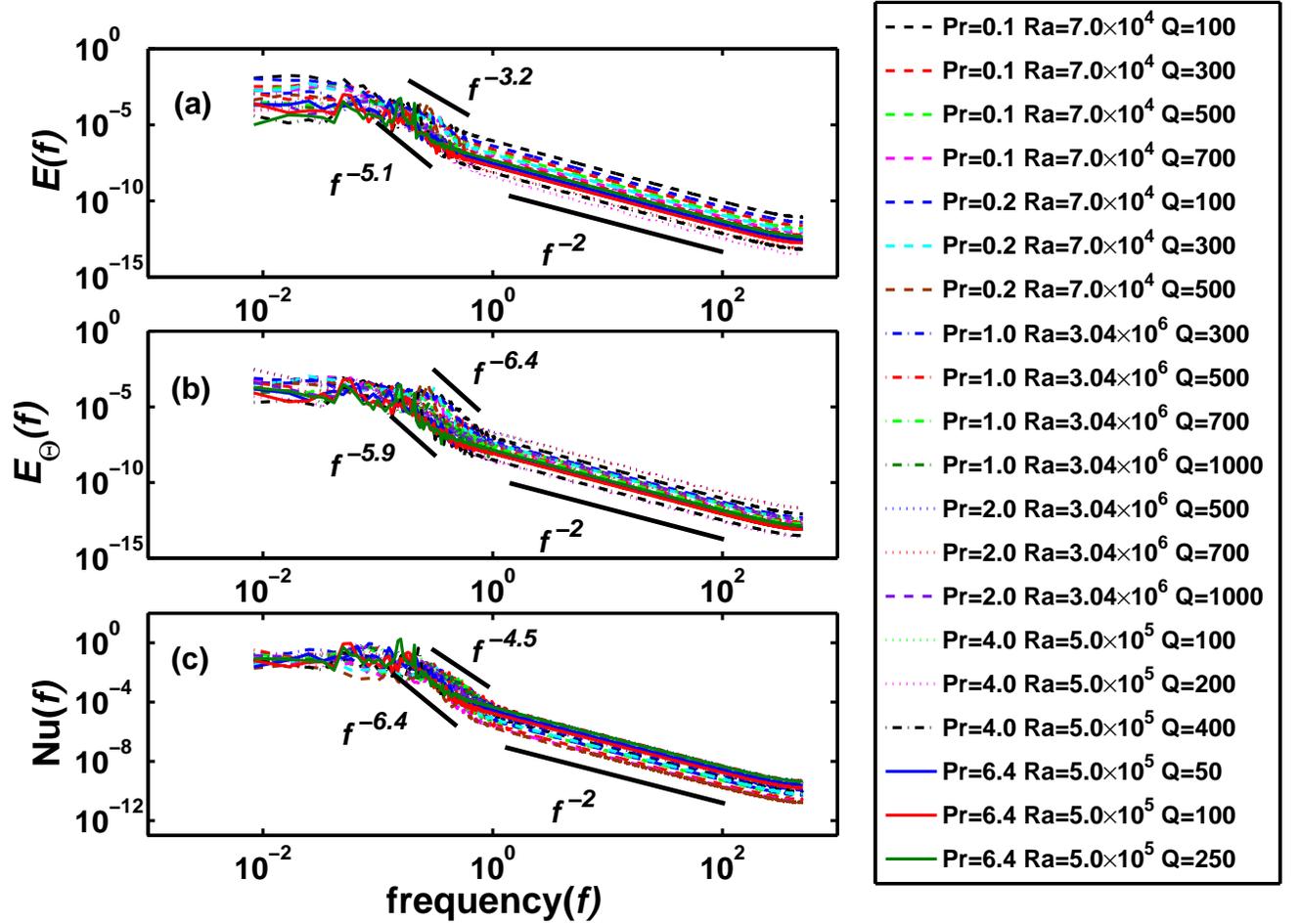}
	\caption{\label{psds} 
		Frequency power spectral densities (PSD) of the energy per unit mass $E(f)=|\mathrm{v}(f)|^2$,  the convective entropy per unit mass $E_{\Theta} (f) = |\theta (f)|^2$ and thermal flux $\mathrm{Nu}(f)$ in the frequency space for different values of $\mathrm{Ra}$, $\mathrm{Q}$ and $\mathrm{Pr}.$  
	}
\end{figure*}

\section{\label{sec:DNS}Direct Numerical Simulations}
\noindent
The direct numerical simulations are carried out using pseudo-spectral method. The perturbative fields are expanded as: 

\begin{eqnarray}
{\bm{\Psi}} (x,y,z,t) &=& \sum_{l,m,n} {\bm \Psi}_{lmn}(t) e^{ik(lx+my)}\cos{(n\pi z)},\label{eq.psi}\\
{\bm{\Phi}} (x,y,z,t) &=& \sum_{l,m,n} {\bm \Phi}_{lmn}(t) e^{ik(lx+my)}\sin{(n\pi z)},\label{eq.phi}
\end{eqnarray}
where ${\bm{\Psi}} (x, y, z, t) = [{\mathrm{v}_1}, {\mathrm{v}_2} , {p}]^{\dagger}$ and ${\bm{\Phi}} (x, y, z, t) = [{\mathrm{v}_3}, {\theta} ]^{\dagger}$. The time dependent Fourier amplitudes of these fields are denoted by ${\bm \Psi}_{lmn} (t) = [U_{lmn}, V_{lmn}, P_{lmn}]^{\dagger}$ 
and ${\bm \Phi}_{lmn} (t) = [W_{lmn}, \Theta_{lmn}]^{\dagger}$, where $l$, $m$ and $n$ are integers. The horizontal wave vector of the perturbative fields is $\bm{k} = lk\bm{\mathrm{e}}_1 + mk\bm{\mathrm{e}}_2$, where $\bm{\mathrm{e}}_1$ and $\bm{\mathrm{e}}_2$ are the unit vectors along the $x$- and $y$-axes. The numerical simulations are carried out in a three dimensional periodic box of size $L\times L\times 1$, where $L=2\pi/k_c(\mathrm{Q})$. The possible values of the integers $l, m, n$  are decided by the continuity equations. They can take values which satisfy the following equation.
  
\begin{equation}
ilk_c (\mathrm{Q}) U_{lmn} + imk_c(\mathrm{Q}) V_{lmn} + n\pi W_{lmn}=0.
\end{equation}
A minimum spatial grid resolution of $128\times 128\times 128$ or $256\times 256\times 256$ has been used for the simulations presented here. The integration in time is performed using a standard fourth order Runge-Kutta  (RK4) method. The data points of the temporal signal  are recorded at equal time  interval of 0.001 to determine record the signals. The time steps have been chosen such that the Courant-Friedrichs-Lewy (CFL) condition is satisfied for all times.


\section{\label{sec:Result}Results and Discussions}

\begin{table*}[ht]
                                                                                                 \begin{ruledtabular}
\def~{\hphantom{0}}
 \begin{tabular}{cccccccccc}	
$\mathrm{Pr}$  & $\mathrm{Ra}$ & $\mathrm{Q}$ & {Exponent $\alpha$}& {Exponent $\beta$} & {Exponent $\gamma$}\\
\hline
$0.1$ & $7.0\times10^4$  & $100$ & $1.97$   & $1.97$ & $1.96 $\\  
		
 &  & $300$ & $1.97$   & $1.97$ & $1.97$\\  
		
 &  & $500$ & $1.96$   & $1.96$ & $1.97 $\\  
		
 &  & $700$ & $1.96$   & $1.97$ & $1.97$\\ \\
		
$0.2$ & $7.0\times10^4$  & $100$ & $1.96$   & $1.97$ & $1.96$\\  
 &  & $300$ & $1.97$   & $1.97$ & $1.96$\\  
		
 &  & $500$ & $1.96$   & $1.96$ & $1.96 $\\ \\ 
		
$1.0$ & $3.04\times10^6$  & $300$ & $1.96$   & $1.97$ & $1.96 $\\  
 &  & $500$ & $1.96$   & $1.97$ & $1.96$\\ 
		
 &  & $700$ & $1.97$   & $1.96$ & $1.96$\\  
		
 &  & $1000$ & $1.96$   & $1.97$ & $1.97$\\  \\
		
$2.0$ & $3.04\times10^6$  & $500$ & $1.96$   & $1.96$ & $1.96$\\  
 &  & $700$ & $1.96$   & $1.96$ & $1.96$\\  
		
 &  & $1000$ & $1.96$   & $1.97$ & $1.96$\\  \\
		
$4.0$ & $5.0\times10^5$  & $100$ & $1.96$   & $1.97$ & $1.96 $\\  
 &  & $200$ & $1.96$   & $1.97$ & $1.97$\\ 
		
 &  & $400$ & $1.96$   & $1.97$ & $1.97$\\  \\
		
$6.4$ & $5.0\times10^5$  & $50$ & $1.96$   & $1.97$ & $1.96 $\\   
 &  & $100$ & $1.97$   & $1.97$ & $1.96$\\ 
 &  & $250$ & $1.97$   & $1.96$ & $1.97$\\
\end{tabular}
\caption{ \label{slopes}
List of Prandtl number $\mathrm{Pr}$, Chandrasekhar number $\mathrm{Q}$, Rayleigh number $\mathrm{Ra}$, exponents of Kinetic  energy$( \alpha)$,exponents of Entropy$( \beta)$ and exponents of Nusselt number $( \gamma)$.}
\end{ruledtabular}
\end{table*}
\begin{figure}[htp]
	\centering
	\includegraphics[height=!,width=9.0 cm]{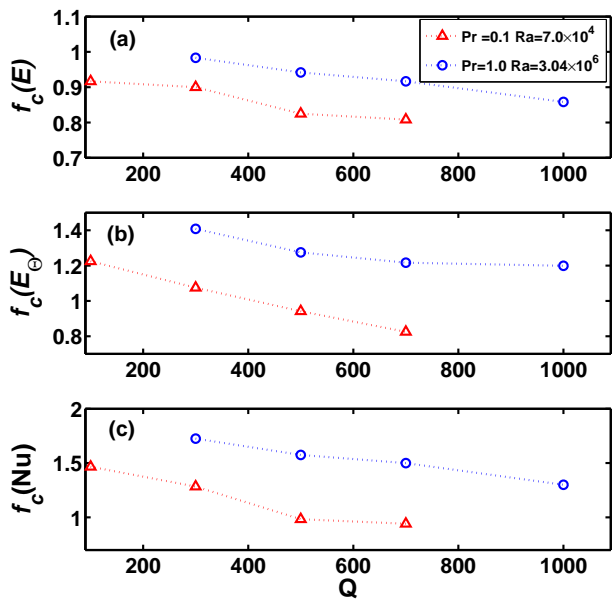}
	\caption{\label{critical_freq} 
		Variation of critical values of the dimensionless frequencies $f_c(E)$, 
		$f_c (E_{\Theta})$ and $f_c (\mathrm{Nu})$ for the energy spectra $E(f)$,  entropy spectra $E_{\Theta}(f)$ and thermal flux $\mathrm{Nu}(f)$, respectively, with  the Chandrasekhar number $\mathrm{Q}$ for  Prandtl number $\mathrm{Pr} = 0.1$ [red(light gray) triangles] and $1.0$ [blue(gray) circles].}
\end{figure}
The simulations are done for several values of thermal Prandtl number ($ 0.1 \le\mathrm{Pr} \le 6.4$). These values of $\mathrm{Pr}$ are relevant for Earth's liquid outer core~\cite{Olson_Glatzmaier_1996}. They are also relevant for problem of crystal growth~\cite{Lan_Kou_1991} and water based nano-fluids~\cite{Kakac_Pramuanjaroenkij_2009}. The Rayleigh number is varied in a range $7.0 \times 10^4 \le \mathrm{Ra} \le 3.04 \times 10^6$, while the Chandrasekhar's number is varied in a range $50 \le \mathrm{Q} \le 10^3$.
Fig~\ref{Time_sig} shows the  temporal variations of three global quantities for $\mathrm{Ra} = 5.0 \times 10^5$, $\mathrm{Pr} = 4.0$ and for two different values of $\mathrm{Q}$: (1) the kinetic energy per unit mass $E$, (2) the convective entropy per unit mass $E_{\Theta}$  and (3) the Nusselt number $\mathrm{Nu}$. All global quantities are averaged over a three-dimensional simulation box described above. The first two set of curves (from the top) show the variations of $E$ with dimensionless time. The light gray (red) curve is for $\mathrm{Q}=100$  and the gray (blue) curve is for  $\mathrm{Q}=400$. The mean of the kinetic energy decrease with increase in $\mathrm{Q}$. The fluctuations of the energy signal also decreases with increase in $\mathrm{Q}$. The curves in  the third and fourth rows from the top show the temporal variations of $E_{\Theta}$, and the curves in the fifth and sixth rows (from the top) show the temporal signal for the Nusselt number $\mathrm{Nu}$, which is a measure of the heat flux. The mean values of the entropy per unit mass and the Nusselt number also decrease with increase in $\mathrm{Q}$. The fluctuations in their temporal signals also decrease with increase in $\mathrm{Q}$. 

Figure~~\ref{psds} displays the power spectrum densities (PSD) for the spatially averaged global quantities in the frequency space for several values of $\mathrm{Ra}$, $\mathrm{Pr}$ and $\mathrm{Q}$. The PSDs of the fluid speed $E(f) = |\mathrm{v}(f)|^2$ are shown in Fig~\ref{psds}(a). The energy spectra are very noisy for dimensionless frequencies between $0.04$ and $1.0$. In this   frequency range ($0.04 < f < 1.0 $), the spectra is noisy and the slope of the curves ${E(f)}-f$ on the log-log scale varies between $-3.2$ to $-5.1$.  However, the $E(f)$ is found to have negligible noise $ 1 < f < 200$. The PSD shows a very clear scaling behaviour for $f > 1$.  The PSD ($E(f)$) of the energy signal  scales with frequency $f$ as almost $f^{-\alpha}$ with $\alpha \approx 2$. The scaling behaviour is found to be continued for more than two decades. The scaling exponent is independent of $\mathrm{Pr}$, $\mathrm{Ra}$ and $\mathrm{Q}$ in this frequency window. Table-I gives the exact values of the exponent $\alpha$ for different values of $\mathrm{Ra}$, $\mathrm{Pr}$ and $\mathrm{Q}$. The scaling law $E(f) \sim f^{-2}$ was also observed in rotating Rayleigh-B\'{e}nard convection (RBC)~\cite{Hirdesh_etal_2014}.
 
Fig~\ref{psds}(b) shows the PSDs of the convective entropy  $E_{\Theta} (f) = |{\theta(f)}|^2$ of the fluid in the frequency space for different values of $\mathrm{Ra}$, $\mathrm{Pr}$ and $\mathrm{Q}$. Its power spectra is also noisy in the dimensionless frequency range $0.04 < f < 1.0 $. The slope on the log-log scale varies between $-5.9$ and $-6.4$. However for $f > 1.0$, $E_{\Theta}$ also scales with frequency with as $f^{-\beta}$ with $\beta \approx 2$. The numerically computed values of the exponent $\beta$ are listed in Table-I. Interestingly, the power spectra of the temperature fluctuations are also found to vary as ${f^{-2}}$ in the  turbulent RBC experiments~\cite{Boubnov-Golitsyn_1990}.

The PSDs for the thermal flux [Nusselt number, $\mathrm{Nu} (f)]$ for several values of  values of $\mathrm{Ra}$, $\mathrm{Pr}$ and $\mathrm{Q}$ are shown in Fig.~\ref{psds}(c). The PSDs also show the scaling behaviour. The PSDs are noisy, as in the case of energy and entropy signals, for dimensionless frequencies  $0.04 < f < 1.0 $. The scaling exponent varies between $-4.5$ to $-6.4$ in this frequency range. However, for dimensionless frequencies range $1 < f < 200$, the spectra for thermal flux $\mathrm{Nu} (f)$ also shows very clear scaling: $\mathrm{Nu}(f)\sim f^{-\gamma}$, where $\gamma \approx 2$. The Table-I shows the values of the exponent $\gamma$ computed in DNS. The measurements of the spectra of thermal flux in RBC also shows the similar scaling law~\cite{Aumaitre_Fauve_epl_2003}.

The scaling law showing the variation of the power spectra as $f^{-2}$ starts at a critical frequency $f_c$ for different values of the Chandrasekhar number. Fig.~\ref{critical_freq} shows the variation of the critical frequency for $E(f)$, $E_{\Theta}(f)$ and $\mathrm{Nu}(f)$ with $\mathrm{Q}$ two different values of $\mathrm{Pr}$. The critical frequency $f_c (E)$ becomes lower as $\mathrm{Q}$ is increased (see Fig.~\ref{critical_freq} (a)). In addition, it is less for smaller values of $\mathrm{Pr}$. Figs.~\ref{critical_freq} (b)-(c) show the variation of $f_c (E_{\Theta})$ and $f_c (\mathrm{Nu})$, respectively, with $\mathrm{Q}$. The values of critical frequencies are slightly different for $E (f)$, $E_{\Theta} (f)$ and $\mathrm{Nu} (f)$. However they all decrease with increase in $\mathrm{Q}$. They also decrease with decrease in the value of $\mathrm{Pr}$.    
\section{\label{sec:conclusion}Conclusions}
Results of direct numerical simulations on Rayleigh-B\'{e}nard magnetoconvection show that power spectral densities the kinetic energy $E(f)$, convective entropy $E_{\Theta}(f)$ and the Nusselt number $\mathrm{Nu} (f)$ scale as $f^{-2}$ for frequencies above a  critical value $f_c$. The critical values $f_c (E)$, $f_c (E_{\Theta})$ and 
$f_c (\mathrm{Nu})$ are different for kinetic energy, convective entropy and the Nusselt number. The critical frequency decreases with increase in the strength of the external magnetic field.   However, the scaling exponent is independent of the thermal Prandtl number, Rayleigh number and Chandrasekhar number. The results may be relevant for geophysical problems, water based nano-fluids and crystal growth.

\section{References}
\noindent

\nocite{*}
\bibliography{aipsamp}

\end{document}